\documentclass[aip,rsi,reprint]{revtex4-1}

\draft 

\usepackage{amssymb,amsmath}
\usepackage{graphicx} 
\usepackage{color}			
\usepackage{textcomp}

\begin{document}

\title{Real-time calibration of a feedback trap} 

\author{Mom\v cilo Gavrilov}
\noaffiliation

\author{Yonggun Jun}
\altaffiliation{Present address:  Department of Developmental and Cell Biology, University of California, Irvine, CA  92697-2300, USA}
\noaffiliation

\author{John Bechhoefer}
\email[email: ]{johnb@sfu.ca}
\noaffiliation

\affiliation{Department of Physics, Simon Fraser University, Burnaby, B.C., V5A 1S6, Canada}

\date{\today}

\begin{abstract}
Feedback traps use closed-loop control to trap or manipulate small particles and molecules in solution.  They have been applied to the measurement of physical and chemical properties of particles and to explore fundamental questions in the non-equilibrium statistical mechanics of small systems.  These applications have been hampered by drifts in the electric forces used to manipulate the particles.  Although the drifts are small for measurements on the order of seconds, they dominate on time scales of minutes or slower.  Here, we show that an extended recursive least-squares (RLS) parameter-estimation algorithm can allow real-time measurement and control of electric and stochastic forces over time scales of hours.  Simulations show that the extended-RLS algorithm recovers known parameters accurately.  Experimental estimates of diffusion coefficients are also consistent with expected physical properties.
\end{abstract}


\maketitle 

\section{Introduction}
\label{sec:intro}

The feedback trap is a new experimental technique developed in 2005 by  Cohen and Moerner, who termed it the ABEL (Anti-Brownian ELectrokinetic) trap.\cite{Cohen2005}  The technique allows trapping of small particles and molecules in solution by creating a feedback loop\cite{bechhoefer2005} where one images the object to be trapped and then applies an electric force to the charged object that moves it in the desired direction.  Because direct electric forces are stronger than the dipolar forces used in optical and magnetic tweezers, feedback traps can trap molecules and submicron particles that are otherwise impossible to confine.  It is even possible to trap a single fluorescent dye molecule in water.\cite{Fields2011,Wang2014}

The feedback trap has had two major types of application.  First, it has been used to probe the physical or chemical properties of the trapped object. These include the diffusion constant and electric-field mobility of single particles,\cite{Cohen2006} their fluctuations,\cite{Cohen2007a} and elastic and dissipation parameters that characterize the internal degrees of freedom of more complex objects.\cite{Cohen2007,Goldsmith2010}  Feedback traps can also be used to estimate the chemical properties of single molecules, including photodynamic and enzymatic properties of biomolecules\cite{Wang2012} and the interplay between fluorescence spectroscopy and conformation at the single-molecule level.\cite{Wang2013,Schlau-Cohen2013}

The second kind of application has been to the study fundamental questions in statistical mechanics.  The key feature of such traps is the ability to impose arbitrary, time-dependent virtual potentials on particles.  For example, Cohen studied the motion of a particle in a virtual double-well potential and in radial potentials of the form $U(r) \sim r^n$.\cite{CohenArbPotential05}  

In preliminary work,\cite{Cho2011} we used time-dependent virtual potentials to study  Landauer's principle, \cite{Landauer1961,Berut2012} which relates information erasure to thermodynamic work.  Our initial attempts were frustrated by small but persistent drifts that, over hours or days, led to significant systematic errors in work measurements.  Extensive investigation showed that the drifts were mostly caused by potential offsets that are linked to chemical reactions at the electrodes used to impose the electric field on the particle.  A secondary cause was temperature-dependent offsets of the voltage amplifier used to impose potential differences across the electrodes.  We conclude that such drifts have been present, not only in our work, but in previous experiments, as well, and have an importance that grows with the duration of measurements.  Improving temperature and potential control can reduce somewhat the magnitude of the drifts but cannot make them small enough to be neglected in long experiments.  A recent discussion of feedback-trap calibration has successfully demonstrated accurate parameter measurements over short time scales (up to one minute).\cite{Wang2014}  The focus here is on time scales of hours, or even days.  Such long times are important in tests of stochastic thermodynamics, which depend on high-precision statistics for particle trajectories.

Here, rather than try to eliminate drifts, we continuously measure and correct for them in real time as the experiment runs.  We will see that such techniques can successfully remove the effects of drifts from experimental data that are collected over days.  We also correct for biases that arise because of subtle correlations in the noise that enter because of the structure of the feedback loop.  In related work, we have used the techniques described here to make the highest precision measurements of Landauer's principle achieved to date. \footnote{
Y. Jun, M. Gavrilov, and J. Bechhoefer, in preparation.}

Below, we describe in detail the methods we used to carry out the real-time calibration of the feedback trap.  We begin, in Sec.~\ref{sec:feedbackTrap} by briefly recalling some of the details of our trap, in particular as they relate to the timing of data acquisition.  In Sec.~\ref{sec:ParticleDynamics}, we review the equations of motion that describe rigid particles in a feedback trap and show how to cast the equations in a more convenient and general form, which is required to implement the calibration.  In Sec.~\ref{sec:par-est}, we discuss the principles of ordinary and extended recursive least-squares parameter estimation.  The basic idea is to recast least-squares fits to allow old fit parameters to be updated each time a new data point is taken, rather than redoing the whole fit.  The resulting speedup is important for real-time operation.  In Sec.~\ref{sec:Trap2d}, we generalize the previous discussion from one to two spatial dimensions.  The former is easier to follow, but the latter is what we actually use.  In Sec.~\ref{sec:control}, we describe our control software.  Because the methods we use are real time, they must be integrated with the rest of the experiment, leading to a rather complex control program, whose details are crucial in achieving a successful calibration.  In Sec.~\ref{sec:simulations}, we show that our analysis methods work on simulated data.  That is, we show that if we simulate data with known parameters, then our analysis routines recover the known parameters.  We will see that the task is complicated by correlations in the noise term that bias the inference if not accounted for.  Finally, in Sec.~\ref{sec:Experiment}, we present experimental results.  We see convergence similar to that observed in simulations and, in the case of the diffusion coefficient measurement, argue that the absolute values are in the expected range, as well.  We conclude that the advances described here show sufficient mastery of the experiment to obtain reliable thermodynamic measurements.

\section{Feedback trap}
\label{sec:feedbackTrap}

\begin{figure}[ht!]
	 \includegraphics[width=5cm]{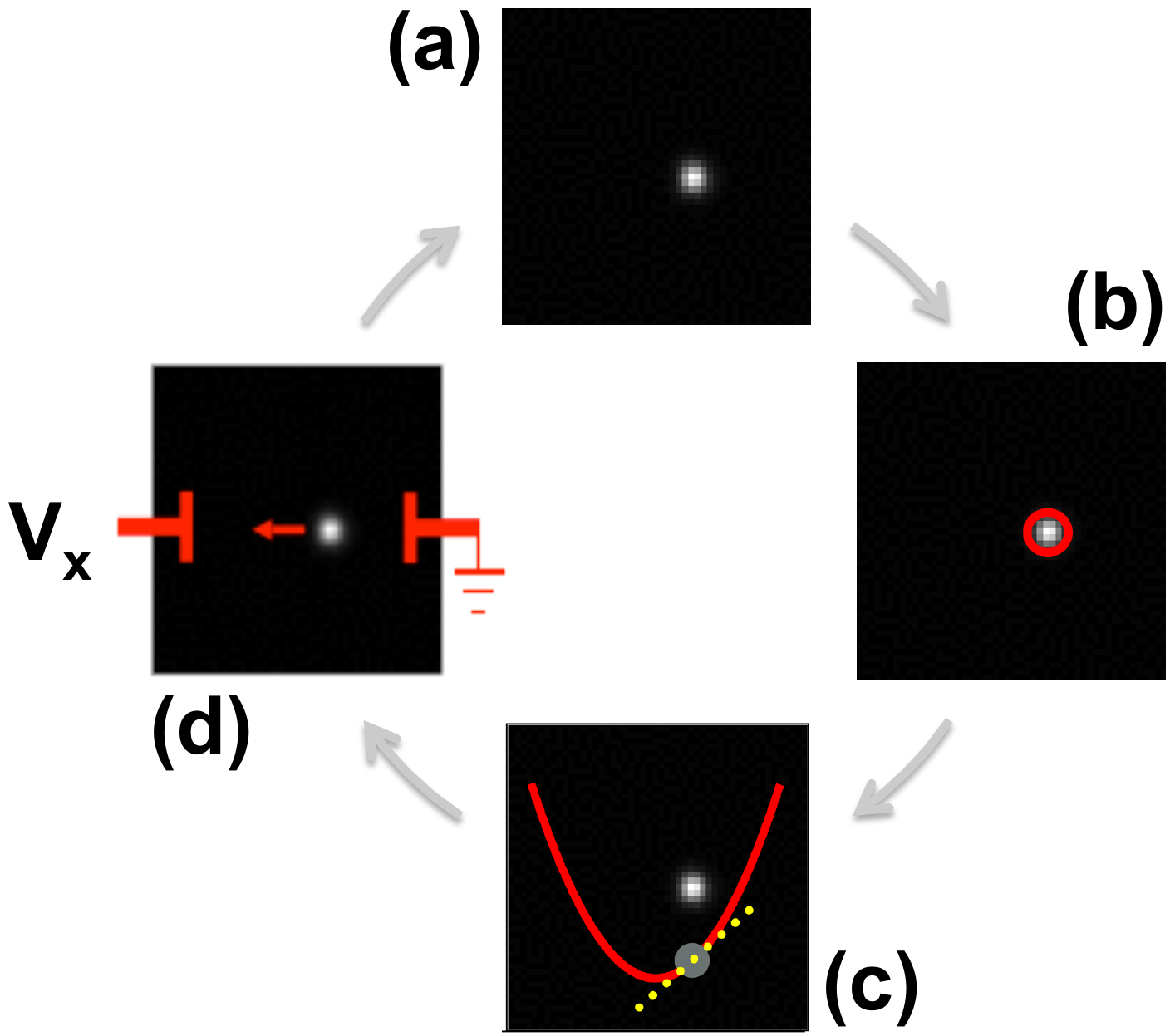}
\caption[example] { \label{fig:FBschematic} 
One time step (one cycle) of a feedback trap:  (a) Acquire an image;  (b) estimate position ($\bar{x}_n$) from the image; (c) calculate feedback force as a gradient of potential  $F_n=-\partial_x U (\bar{x}_n,t)$;  (d) apply feedback force as an electric force via set of electrodes.  At the end of the cycle, deterministic and stochastic forces will have changed the particle position.}
\end{figure}

The operation of a feedback trap is illustrated in Fig.~\ref{fig:FBschematic}.  The trap first acquires an image of an object and uses image-analysis software to estimate its position [Fig.~\ref{fig:FBschematic}(a) and (b)].  Based on the observed position ($\bar{x}_n$) and the chosen virtual potential, the program calculates the force to be applied as the negative gradient of the imposed potential  [Fig.~\ref{fig:FBschematic}(c)].  This force is applied as an electrical force by applying voltage over the set of two horizontal electrodes  given in Fig.~\ref{fig:FBschematic}(d).  At the end of the cycle, a particle has been displaced relative to its previous position because of feedback and thermal (diffusion) forces.  We use a cycle time of 10 ms.

	The scheme in Fig.~\ref{fig:FBschematic} is easily generalized to the two-dimensional case by calculating the position along the other axis and inserting an  additional set of ``vertical" electrodes in Fig.~\ref{fig:FBschematic} (d).  We thus apply forces along two directions independently.

\subsection{Experimental setup}
\label{sec:feedbackTrap}

Our version of the feedback trap, described more completely in Ref.~\onlinecite{Gavrilov2013}, uses 210-nm polystyrene beads in water, confined in the vertical dimension in an 800-nm-thick cell and controlled in the two lateral dimensions by two  sets of electrodes.  Particle images are recorded using an EM-CCD camera mounted to a home-built epifluorescence microscope.  The digital image is loaded into a LabVIEW program that uses a centroid algorithm to determine the particle position, calculate the required forces, and output voltages to two data acquisition devices (DAQs).  The analog output of the first DAQ controls the intensity of the excitation laser, regulating it so that the detected fluorescence intensity stays constant, even as the particle bleaches, thus keeping the measurement noise constant.

The analog outputs of the other DAQ are used to apply forces on the particle.  They are sent to a home-built voltage amplifier with a gain of 15 and then applied to two pairs of electrodes.  To limit current flows through the cell due to capacitative charging, we place a 10 k$\Omega$ resistor in parallel with each pair of electrodes and 1k$\Omega$ resistor in series.  Since the electrical resistance of the flow cell is $\approx 10$ M$\Omega$, the voltage drop due to the series resistor is negligible.

\begin{figure}[ht!]
	 \includegraphics[width=8cm]{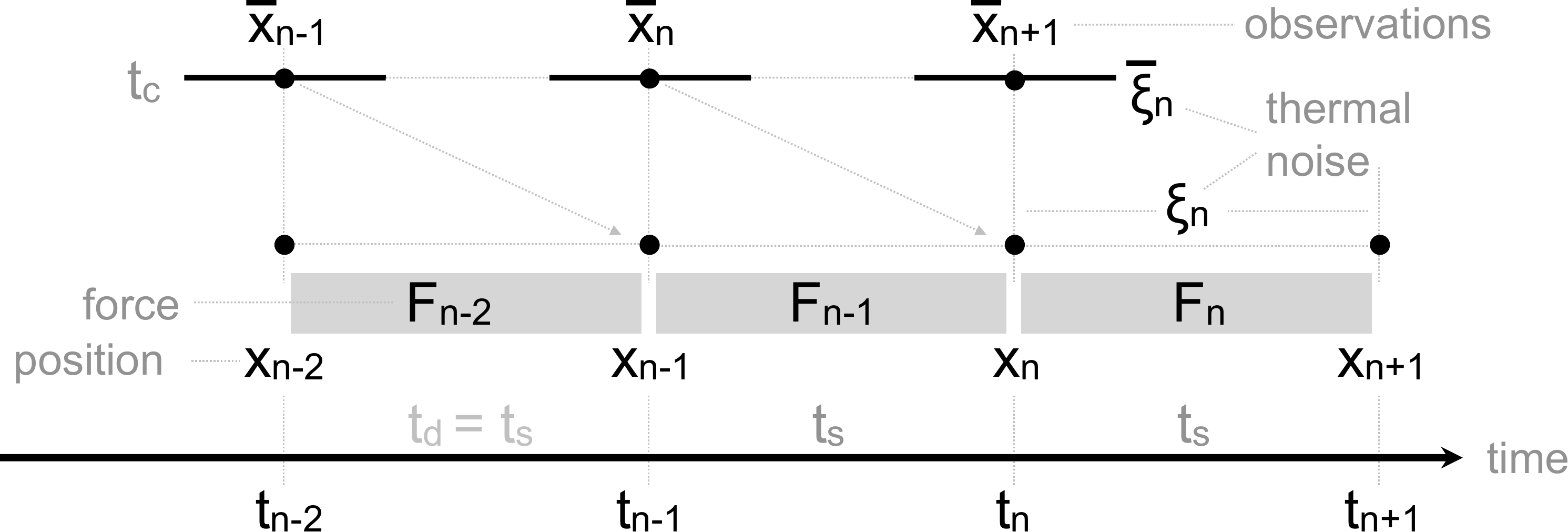}
\caption[example] { \label{fig:Timing} 
Timing diagram for the feedback trap.  The particle's position is averaged over an exposure time $t_c$, and forces are updated after $t_s$.  The observed position $\bar{x}_n$ is used for calculating feedback force $F_n$. The force is applied with a time delay  $t_d=t_s$, measured from the middle of the exposure. Both the deterministic force $F_n$ and the thermal force $\xi_n$ affect the unknown real position $x_{n+1}$.}
\end{figure}

The update time of the feedback trap is set to 10 ms, while the camera exposure time is 5 ms.  The delay between the mid-point of the camera exposure and the application of a feedback voltage response is  set to 10 ms.  The timing diagram for the data acquisition is shown in Fig.~\ref{fig:Timing}.

\section{Particle dynamics in a feedback trap}
\label{sec:ParticleDynamics}

Previously,\cite{Jun2012} we derived the equations of motion for a particle in a one-dimensional virtual harmonic potential of the form $U_{\rm harm}(x) = \tfrac{1}{2} \kappa x^2$, where $\kappa$ is the force constant and $x$ the displacement from equilibrium.  Here, we generalize to the case of motion in an arbitrary, time-dependent virtual potential $U(x,t)$ and then derive an alternate form of the equations of motion that is more convenient for numerical simulations and for inferring parameter values.

\subsection{Coupled one-dimensional equations of motion}
\label{sec:coupled1d}

  Let $F_n$ be the force due to the virtual potential, held constant over the time interval $t_s$, defined by $[t_n, t_{n+1}) \equiv [nt_s,(n+1)t_s)$.  An important point is that the virtual potential is based on the observed position $\bar{x}_n$ rather than the unknown true position $x_n$.  The observed position $\bar{x}_n$ is inferred from a camera image that averages the motion over a time $t_c<t_s$.  The response to each camera exposure is to update the desired force and necessarily involves a finite time delay, $t_d$, which must be at least $\tfrac{1}{2} t_c$.  In our experiments, for simplicity, we set $t_d = t_s$, as illustrated in Fig.~\ref{fig:Timing}.

With that choice of delay, the dynamics are determined by two coupled, discrete stochastic equations for $x_n$ and $\bar{x}_n$.  In the overdamped limit, which applies in all cases we study, the equations are\cite{Jun2012}
\begin{subequations}
\begin{align}
	\label{eq:Langevin_true}
	x_{n+1} &= x_n + \tfrac{1}{\gamma}F_nt_s + \xi _n  \\ 
	\label{eq:Langevin_observed}
	\bar{x}_{n+1} &= x_n+\tfrac{1}{8\gamma}(F_n-F_{n-1})t_c - \xi _n^{(0)}
		+\bar{\xi}_n + \chi_n \,.
\end{align}
\label{eq:Langevin_general}
\end{subequations}
In Eq.~\ref{eq:Langevin_true}, the true position $x_{n+1}$ depends on the deterministic force $F_n = -\partial_x U(\bar{x}_n, t_n)$ applied during time $t_s$ and on the thermal noise $\xi_n$.  The particle drag coefficient $\gamma$ gives the response to the feedback force and is related to the diffusion coefficient by Einstein's law, $D=kT/\gamma$.  As we discuss below in Sec.~\ref{sec:diffusion}, the confined vertical geometry implies that the drag coefficient $\gamma$ is significantly larger than the Stokes-Einstein value for an isolated sphere.  The effects of thermal fluctuations are approximated, as usual, by a Gaussian random variable $\xi_n$, which satisfies $\langle \xi_n \rangle = 0$ and $\langle \xi_n \, \xi_m\rangle = 2Dt_s \, \delta _{mn} $, where $\delta_{mn}$ is the Kronecker delta symbol.

In Eq.~\ref{eq:Langevin_observed}, the observed position $\bar{x}_{n+1}$ depends on the true position $x_n$ but also on the feedback forces during previous time steps, via the term $\tfrac{1}{8\gamma}(F_n-F_{n-1})t_c$.  This term results from the finite camera exposure: the timing of the feedback trap is set so that the force is updated, from $F_{n-1}$ to $F_n$, at  precisely the midpoint of the camera exposure.  A change in force at the update then biases the position, accounting for the $\tfrac{1}{8\gamma}(F_n-F_{n-1})t_c$ term.  The noise terms include the effects of thermal noise as averaged over the camera exposure $t_c$, and the observation noise $\chi_n$.  A detailed explanation and derivation of Eq.~\ref{eq:Langevin_general} is given in Ref.~\onlinecite{Jun2012}.

\subsection{Reduction to a single equation of motion}
\label{sec:reduce-to-single-equation}

It is convenient to eliminate the unobservable true position $x_n$ from Eq.~\ref{eq:Langevin_general}, reducing the two coupled equations to a single relation that is more suitable for simulation and further analysis.  To accomplish this reduction, we  convert absolute positions into intervals, defining $\Delta x_n \equiv x_{n+1}-x_n$ for true positions and 
$\overline{\Delta x}_n \equiv \bar{x}_{n+1}-\bar{x}_{n}$ for observed positions.  Then,\begin{widetext}
\begin{equation}
	\overline{\Delta x}_n = \Delta x_{n-1} + \tfrac{1}{8\gamma}(F_n-2F_{n-1}
		+ F_{n-2})t_c  - \xi _n^{(0)} + 
		\xi _{n-1}^{(0)}+\bar{\xi}_n-\bar{\xi}_{n-1}+ \chi_n- \chi_{n-1} \,.
\label{eq:LangevinCompact1}
\end{equation}
Substituting $\Delta x_{n-1} = \tfrac{1}{\gamma} F_{n-1}t_s + \xi_{n-1}$ from Eq.~\ref{eq:Langevin_true} gives
\begin{align}
	\overline{\Delta x}_n &= \frac{t_s}{\gamma} \underbrace{\left[ F_{n-1}
	+ \tfrac{1}{8} \, \tfrac{t_c}{t_s} \, (F_n-2F_{n-1}+F_{n-2}) 
		\right]}_{\text{deterministic forces } \equiv \bar{F}_{n-1}} 
	+ \underbrace{ \xi_{n-1}- \xi _n^{(0)}+ \xi _{n-1}^{(0)}
	+ \bar{\xi}_n-\bar{\xi}_{n-1}+ \chi_n- \chi_{n-1}}_{\text{stochastic terms } \equiv \zeta_n} 
	\equiv  \frac{t_s}{\gamma} \bar{F}_{n-1} + \zeta_n \,, 
\label{eq:LangevinCompact2}
\end{align}
\end{widetext}
where terms proportional to $t_s/\gamma$ are deterministic and are collected into a single effective force, $\bar{F}_{n-1}$.  The camera-exposure corrections are small (typically $\lesssim 6\%$) so that $\overline{F}_{n-1} \approx F_{n-1}$.  See Appendix~\ref{sec:camera-exposure} for a discussion.

The terms forming $\zeta_n$ in the second grouping are stochastic and result from thermal fluctuations and observation noise.  The camera integrates Brownian motion of the particle over the  exposure time $t_c$, while the observation noise $\chi_n$ results from the finite resolution of the optical microscope and the finite number of photons collected during the camera exposure.  

Since $\zeta_n$ is a linear combination of Gaussian random variables with zero mean, it, too, has zero mean and is characterized solely by its covariance.  The effective noise $\zeta_n$ is not white but has correlations\cite{Goulian2000,Savin2005,cohen06c,Jun2012}
\begin{align}
	\langle \zeta_n^2\rangle &= 2Dt_s-\tfrac{2}{3}Dt_c+2\chi^2 \nonumber \\
	\langle \zeta_n \, \zeta_{n-1}\rangle &= \tfrac{1}{3}Dt_c-\chi^2 \nonumber \\
	\langle \zeta_n \, \zeta_{n-p}\rangle &= 0 \,, \qquad \qquad p \ge 2 \,.
\label{eq:Zeta-corr}
\end{align}
The cross-correlation $\langle \zeta_n \, \zeta_{n-1}\rangle$ arises because neighboring intervals,  $\overline{\Delta x}_n = \bar{x}_{n+1}-\bar{x}_n$ and $\overline{\Delta x}_{n-1} = \bar{x}_n-\bar{x}_{n-1}$ share the noise that is present in measurement $\bar{x}_n$.  Equation~\ref{eq:Zeta-corr} shows that nearest-neighbor correlations can be positive or negative, or even zero with a judicious choice of $t_c$.  Note that the observation  noise variance $\chi^2$ is shot-noise limited and thus $\sim t_c^{-1}$.  

For simulations, it is convenient to write the correlated noise term $\zeta_n$ in terms of uncorrelated Gaussian random variables $\psi_n$, with $\langle \psi_n \rangle = 0$ and $\langle\psi_n\psi_m\rangle=\delta_{nm}$.   More succinctly, $\psi_n \sim \mathcal{N}(0,1)$, and  
\begin{align}
	\zeta_n &\equiv c_+\psi_n+ c_-\psi_{n-1} \,, \nonumber \\
	c_\pm &= \frac{1}{2}\left( \sqrt[]{2Dt_s}
		\pm \sqrt{2Dt_s-\tfrac{4}{3}Dt_c+4\chi^2} \right)\,,
\label{eq:zetaCD} 
\end{align}
where the constants $c_\pm$ are chosen to satisfy Eq.~\ref{eq:Zeta-corr}.

\subsection{Generating feedback voltages}
\label{sec:volts1d}

Equation~\ref{eq:LangevinCompact2} does not fully specify the equations of motion, since we still need to relate the forces $F_n$ to the applied voltages $V_n$ and to the desired virtual potential $U(x,t)$.  We first relate the desired forces to voltages placed across the electrodes in the feedback trap.  For the range of applied voltages (generally, $|V_n| <10$~V), the response is linear, so that we expect $F_n = \tfrac{\mu}{\gamma} V_n$, where $V_n$ is the voltage applied across the electrodes, $\gamma$ is the particle drag, and $\mu$ is the mobility.  Empirically, however, we find that $F_n = \tfrac{\mu}{\gamma}(V_n - V_{0})$,  where the voltage $V_{0}$ leads to a drift and reflects amplifier and electrochemical offsets in the applied voltage.  The drift is important on time scales of minutes and longer.  Substituting this relation into the expression for $\bar{F}_{n-1}$ in Eq.~\ref{eq:LangevinCompact2} shows that $\bar{F}_{n-1} = \tfrac{\mu}{\gamma}(\bar{V}_{n-1} - V_{0})$, where
\begin{equation}
	\bar{V}_{n-1} = V_{n-1} + \tfrac{1}{8} \, \tfrac{t_c}{t_s} \, (V_n-2V_{n-1}+V_{n-2}) \,.
\label{eq:Veff}
\end{equation}
The equation of motion then becomes
\begin{equation}
	\overline{\Delta x}_n = t_s\mu(\bar{V}_{n-1}-V_{0}) + \zeta_n\,.
\label{eq:dynamicsV}
\end{equation}
We note that ``mobility" is not quite the correct terminology, as the standard definition relates the particle drift velocity to the local field; however, since determining fields is difficult, it is common practice to state a mobility in terms of the potential drop across the electrodes.\cite{Wang2014}

We next need to relate forces to the desired virtual potential $U(x,t)$.  At time $t_n = n t_s$, we output the voltage $V_n$, which is based on the measurement $\bar{x}_n$ (whose midpoint is at $t_{n-1}$, as shown in Fig.~\ref{fig:Timing}).  We choose this voltage so that 
\begin{equation}
	\tfrac{\mu}{\gamma}(V_{n-1} - V_{0}) = -\partial_{x}U(\bar{x}_{n-1},t_{n-1}) \,.
\label{eq:Force1}
\end{equation}
In principle, we should use $\bar{V}_{n-1}$ in Eq.~\ref{eq:Force1}; however, that choice leads to a numerically unstable algorithm because it involves taking the ratio of two small numbers, $\tfrac{t_c}{8t_s}$ and $[\partial_x U(x_{n-1},t_{n-1}) - \tfrac{\mu}{\gamma}V_{n-1}]$.  Thus, we take advantage of the fact that $\bar{V}_n \approx V_n$ in writing Eq.~\ref{eq:Force1}.  (See Appendix~\ref{sec:camera-exposure}.)  Explicitly, 
\begin{equation}
	V_{n-1} = - \mu^{-1} \partial_x U(x_{n-1}, \, t_{n-1})\gamma  + V_{0} \,.
\label{eq:ApproxDx}
\end{equation}
A final simplification is to redefine the drag $\gamma$ in terms of diffusion using Einstein's relation, $\gamma = (kT)/D$ and to write the potential in units of $kT$.  Then, collecting all the equations together, we have
\begin{align}
	\overline{\Delta x}_n &= t_s\mu(\bar{V}_{n-1}-V_{0}) + \zeta_n \nonumber \\
	\bar{V}_{n-1} &\equiv 
		V_{n-1}+\left( \tfrac{1}{8} \right) \, \left( \tfrac{t_c}{t_s} \right) \, 
		(V_n - 2V_{n-1} + V_{n-2})  \nonumber \\ 
	V_{n} &= - \mu^{-1} \partial_x U(x_{n}, \, t_{n})/D  + V_{0} \nonumber \\
	\zeta_n &= c_+\psi_n+ c_-\psi_{n-1} \nonumber \\
	c_\pm &= \frac{1}{2}\left( \sqrt[]{2Dt_s}
		\pm \sqrt{2Dt_s-\tfrac{4}{3}Dt_c+4\chi^2} \right) \nonumber \\
	\psi_n &\sim \mathcal{N}(0,1) \,.
\label{eq:summary}
\end{align}
Finally, we note that Eqs.~\ref{eq:summary} assume that, over the timescale $t_s$, the equivalent continuous potential does not change significantly.  Such changes can occur in two ways:  by the motion of the particle in a fixed potential and by the time-dependence of the potential itself.  For the former, we ask that $\alpha \equiv t_s/t_r \ll 1$, where $t_r$ is the relaxation time for motion in a potential.  In a harmonic potential with force constant $\kappa$, the relaxation time would be $t_r = \sqrt{\gamma/\kappa}$.  Here, with a general $U(x,t)$, the force constant generalizes to $\kappa = -\partial_{xx} U(x,t)$, which is approximately the curvature of the potential function.  Conservatively, $\kappa(x,t)$ should be evaluated at the point of maximum curvature.  Note that Eq.~\ref{eq:Force1} is for a single force $F_{n-1}$ and must be generalized to $\bar{F}_{n-1}$ using the definition in Eq.~\ref{eq:LangevinCompact2}.

\section{Online parameter estimation}
\label{sec:par-est}

Equation~\ref{eq:summary} contains four undetermined parameters:  the mobility $\mu$,  drift $V_0$, and noise terms $c_\pm$.  From $c_\pm$, we can deduce $D$ and $\chi$, assuming that $t_s$ and $t_c$ are known.  (The latter are known, either because the  hardware gives deterministic control over timing or because we measure $t_s$ and $t_c$ independently, as described in Ref.~\onlinecite{Gavrilov2013}.)  As discussed in the Introduction, we need to measure the  parameters experimentally in real time, while the experiment is running, as opposed to off-line analysis of recorded data.  Real-time parameter values are required in order to impose correctly the proper virtual potential.  In particular,
\begin{itemize}
\item $\mu$  relates displacements to voltages;
\item $V_0$  allows drift compensation;
\item $c_{\pm}$ sets the scale of the potential relative to $kT$.
\end{itemize}

Since most of our experiments last several days and since photobleaching limits the particle lifetime in a feedback trap to a few hours, we need to acquire data for several different particles.  Each particle has its own radius and charge, which translates to an individual diffusion coefficient $D$ and mobility $\mu$, which must be estimated.  The mobility also varies significantly with location in the cell.  The voltage offset $|V_0|$ is typically $\approx 200$ mV and the mobility $\approx 10 \, \mu$m/s/V.  Together, these lead to drift velocities $v_0 = \mu V_0$ that are typically $2 \, \mu$m/s.  The drift terms become comparable to diffusion on time scales of roughly a second, with $v_0 t^* \sim \sqrt{Dt^*} \implies t^* \sim D/v_0^2 \approx$ 1 s.  (These are worst-case estimates; often $t^* \approx 30$ s.)  We thus need an algorithm that can calculate a running average of the parameter estimates over many time steps while still being fast enough to update at each time step, since at each time step we need to output the correct force, based on the current calibration.  The RLS algorithm described in the next section can fulfill both requirements.

\subsection{Recursive Least Squares (RLS)}
\label{sec:RLS}

We begin by formulating the parameter-estimation problem as a  linear, least-squares curve-fit.  We first rewrite Eq.~\ref{eq:summary} in a vector form:
\begin{equation}
	\overline{\Delta x}_n = \boldsymbol{\varphi}_n^T\boldsymbol{\theta} + \zeta_n\,, 
\label{eq:RLStheory}
\end{equation}
where $\boldsymbol{\varphi}^T_n = \bigl( \begin{smallmatrix} \bar{V}_{n-1} && 1 \end{smallmatrix} \bigr) $ and $\boldsymbol{\theta} =t_s\mu\bigl( \begin{smallmatrix} 1 \\ -V_{0} \end{smallmatrix} \bigr)$.  If we neglect, for now, correlations in the noise term $\zeta_n$, the optimal estimate $\hat{\boldsymbol{\theta}}$ of the parameters is determined by minimizing the function \begin{equation}
	\chi^2(\boldsymbol{\theta}) = \sum^N_{n=0} (\overline{\Delta x}_n 
		- \boldsymbol{\varphi}_n^T \boldsymbol{\theta})^2 \,.
\label{eq:chi2}
\end{equation}

%

Because parameter values drift, we will need to estimate them at every time step.  Although in principle one could minimize $\chi^2(\boldsymbol{\theta})$ at each time step, it is well known that the least-squares problem can be formulated \textit{recursively}, with updated estimates of the parameters inferred from old estimates and new data [see, e.g., Ref.~\onlinecite{Astrom2008}, Ch. 2].  If we again assume  decorrelated noise, the recursive algorithm is given by iterating the three RLS equations:  
\begin{align}
	\hat{\boldsymbol{\theta}}_{n+1} &= \hat{\boldsymbol{\theta}}_{n} 
		+ \boldsymbol{L}_{n+1} \varepsilon_{n} \,,   \nonumber \\[3pt]
	\boldsymbol{L}_{n+1} &= \frac{\boldsymbol{P}_{n}\boldsymbol{\varphi}_{n+1}}
		{1 + \boldsymbol{\varphi}^T_{n+1}\boldsymbol{P}_n
		\boldsymbol{\varphi}_{n+1}} \,, \nonumber \\[3pt]
	\boldsymbol{P}_{n+1} &= (\boldsymbol{I} -\boldsymbol{L}_{n+1} 
		\boldsymbol{\varphi}^T_{n+1})\boldsymbol{P}_{n} \,,
\label{eq:RLS-const}
\end{align}
where $\boldsymbol{P}_{n+1}$ is the parameter covariance matrix and where $\varepsilon_n = \overline{\Delta x}_{n+1} - \boldsymbol{\varphi}^T_{n+1} \hat{\boldsymbol{\theta}}_{n}$ defines the \textit{innovations}, the difference between  observed and predicted displacements.  The RLS algorithm is a simplified version of the \textit{Kalman filter},\cite{bechhoefer2005} and the vector $\boldsymbol{L}_n$ is known as the \textit{Kalman gain}:  it gives the relative weight of the old parameter estimates and the new information contained in $\varepsilon_n$.  The estimate $\hat{\boldsymbol{\theta}}_{n}$ includes all data collected up to timestep $n$.

\subsection{Decorrelating the noise}
\label{sec:decorr-noise}

If one does not account for correlations in the noise term $\zeta_n$, the parameter estimates will be biased.  We can avoid bias by transforming to new coordinates where the noise terms are independent.\cite{Astrom2008}  We change variables in Eq.~\ref{eq:LangevinCompact2} by first applying the $Z$ transform, or generating function, which is a discrete version of the Laplace transform.  For the sequence $\zeta_n$, the $Z$ transform is
\begin{equation}
	\mathcal{Z}[\zeta_n] = \zeta(z) \equiv 
	\sum_{n=0}^{\infty}\zeta_n \, z^{-n} \,.
\label{eq:forgetting-factor}
\end{equation}
Since $\zeta_n = c_+ \psi_n + c_- \psi_{n-1}$ and $\mathcal{Z}[\psi_{n-1}] = z^{-1} \psi(z)$, Eq.~\ref{eq:RLStheory} implies that 
\begin{equation}
	\Delta x(z) = \left[ \boldsymbol{\varphi}(z) \right]^T \boldsymbol{\theta} 
		+ \left( c_+ + c_- z^{-1} \right) \psi(z) \,,
\label{eq:filter1}
\end{equation}
where $\boldsymbol{\varphi}^T(z) = \bigl( \begin{smallmatrix} \bar{V}(z) && \tfrac{1}{1-z^{-1}} \end{smallmatrix} \bigr)$.  Dividing Eq.~\ref{eq:filter1} by $\left( c_+ + c_- z^{-1} \right)$ then gives
\begin{equation}
	\Delta x^{(f)}(z) = \left[ \boldsymbol{\varphi}^{(f)} (z)\right]^T \,
		\boldsymbol{\theta} + \psi(z) \,,
\label{eq:filter2}
\end{equation}
where the \textit{filtered} versions of $\Delta x(z)$ and $\boldsymbol{\varphi}(z)$ are 
\begin{align}
	\Delta x^{(f)}(z) &= \frac{\Delta x(z)}{c_+ + c_- z^{-1}} \,, \nonumber \\[3pt]
	\boldsymbol{\varphi}^{(f)}(z) &= 
	\begin{pmatrix} \frac{\bar{V}(z)}{c_+ + c_- z^{-1}} &&
		\left( \frac{1}{1-z^{-1}} \right) \, \left( \frac{1}{c_+ + c_- z^{-1}} \right) 
		\end{pmatrix} \,.
\label{eq:filter3}
\end{align}
Multiplying Eq.~\ref{eq:filter3} by $\left( c_+ + c_- z^{-1} \right)$ and inverting the  $Z$ transform gives recursive formulae for the  filtered observed position $\Delta\bar{x}^{(f)}_n$ and voltage term $\boldsymbol{\varphi}^{(f)}_n$:
\begin{align}
	\Delta \bar{x}_{n} &= c_+\Delta \bar{x}^{(f)}_n + c_-\Delta \bar{x}^{(f)}_{n-1} 
		\nonumber \nonumber \\
	&\implies \quad \Delta \bar{x}^{(f)}_n = \tfrac{1}{c_+}
		\left( \Delta \bar{x}_{n}-c_-\Delta \bar{x}^{(f)}_{n-1} \right) \nonumber \\[6pt]
	\boldsymbol{\varphi}_{n} &= c_+\boldsymbol{\varphi}^{(f)}_n + c_-
		\boldsymbol{\varphi}^{(f)}_{n-1} \nonumber \\
	&\implies \quad \boldsymbol{\varphi}^{(f)}_n = \tfrac{1}{c_+}
		\left( \boldsymbol{\varphi}_{n} - c_-\boldsymbol{\varphi}^{(f)}_{n-1} \right) \,.
\label{eq:filter4}
\end{align}
In components, the form of the filtered input in Eq.~\ref{eq:filter4} is $\left( \boldsymbol{\varphi}^{(f)}_n \right)^T = (\bar{V}^{(f)}_{n-1}\, c_0^{-1})$, where $c_0 = c_+ + c_- = \sqrt{2Dt_s}$.  In terms of the filtered variables, the relation between displacement and voltages becomes 
\begin{equation}
	\Delta \bar{x}^{(f)}_n = \left( \boldsymbol{\varphi}^{(f)}_n \right)^T 
		\boldsymbol{\theta} + \psi_n\,, 
\label{eq:Filtered-fit}
\end{equation}
Thus, we first recursively filter $\bar{x}_n$ and $\boldsymbol{\varphi}_n$ and then use the resulting $\bar{x}^{(f)}_n$ and $\varphi^{(f)}_n$ in an ordinary RLS algorithm to estimate $\hat{\boldsymbol{\theta}}_n$.

\subsection{Time-dependent parameters}
\label{sec:time-dependent-pars}

The algorithms for determining $\mu_n$ and $V_{0n}$ given in the previous section implicitly assume that the underlying parameter values are constant.  Empirically, they drift.  To account for the drift, we can reformulate a running-average version of RLS that weights recent observations more than ones taken in the past.\cite{Astrom2008}  For $N$ measurements, we write the exponentially weighted $\chi^2$ function as 
\begin{equation}
	\chi^2= \sum^N_{n=0} \lambda^{N-n}(\overline{\Delta x}_n 
	- \boldsymbol{\varphi}_n^T\hat{\boldsymbol{\theta}}_n)^2 \,,
\label{eq:RLStheoryPar}
\end{equation}
which leads to a slightly altered version of the RLS equations (Eqs.~\ref{eq:RLS-const}):
\begin{align}
	\hat{\boldsymbol{\theta}}_{n+1} &= \hat{\boldsymbol{\theta}}_{n} 
		+ \boldsymbol{L}_{n+1} \varepsilon_{n} \,,   \nonumber \\[3pt]
	\boldsymbol{L}_{n+1} &= \frac{\boldsymbol{P}_{n}\boldsymbol{\varphi}_{n+1}}
		{\lambda+ \boldsymbol{\varphi}^T_{n+1}\boldsymbol{P}_n
		\boldsymbol{\varphi}_{n+1}} \nonumber \\[3pt] 
	\boldsymbol{P}_{n+1} &= \tfrac{1}{\lambda}(\boldsymbol{I} 
		-\boldsymbol{L}_{n+1}\boldsymbol{\varphi}^T_{n+1})\boldsymbol{P}_{n} \,,
\label{eq:RLS-time-var}
\end{align}
In Eq.~\ref{eq:RLS-time-var}, the \textit{forgetting factor} $\lambda \in (0,1)$, with $\lambda=1$ implying that all measurements are equally weighted.  The forgetting factor is conveniently expressed in terms of a timescale as $\lambda = 1-1/\tau$, since $\lambda^n = (1-1/\tau)^n \approx e^{-n/\tau}$, with $n$ an integer and $\tau$ in units of the time step $t_s$.  The forgetting time $\tau$ should be chosen shorter than the drift,  to track parameter variations.

Finally, we estimate the particle diffusion coefficient $D$ and the observation noise $\chi$.  The equations for $c_{\pm}$ relate the noise correlations of $\zeta_n$ to the diffusion constant and observation noise.  From Eq.~\ref{eq:summary},
\begin{equation}
	\zeta_n=\overline{\Delta x}_n - \mu_n t_s (\bar{V}_{n-1} - V_{0 (n-1)}) \,.
\label{eq:zeta_n}
\end{equation}
After obtaining $\zeta_n$, we calculate running averages of the variance and correlation functions:  
\begin{align}
	\langle \zeta^2\rangle_n &= \langle \zeta^2\rangle_{n-1} + 
		\lambda \left( \zeta_n^2 - \langle \zeta^2\rangle_{n-1} \right) \nonumber \\
	\langle \zeta\zeta_-\rangle_n &= \langle \zeta \zeta_- \rangle_{n-1}
		+ \lambda \left[ \left(\zeta \zeta_- \right)_n - \langle \zeta\zeta_-\rangle_{n-1} \right] \,.
\label{eq:DandChi2}
\end{align}
where $\lambda$ again sets the filtering time, $\langle \zeta^2\rangle_n$ and $\langle \zeta\zeta_-\rangle_n$ are estimates of the variance and unit-lag covariance, respectively, and $(\zeta \zeta_-)_n$ is the new lag-one covariance at timestep $n-1$.  The diffusion and the observation noise are then 
\begin{align}
	D_n &=\frac{1}{2t_s} \left( \langle \zeta^2 \rangle_n
		+ 2\langle \zeta \zeta_-\rangle_n \right) \nonumber \\
	\chi^2_n &= \tfrac{1}{3}D_nt_c-\langle \zeta\zeta_-\rangle_n \,.
\label{eq:DandChi1}
\end{align}
The estimators in Eq.~\ref{eq:DandChi1} are optimal for short single-particle trajectories. \cite{vestergaard14}  

Notice that estimating $D$ and $\chi$ (equivalently, $c_\pm$) requires estimates of $\mu$ and $V_0$ (see Eq.~\ref{eq:zeta_n}), while estimates of $\mu$ and $V_0$ depend on the filtering operation to decorrelate the noise and require estimates of $c_\pm$ (see Eq.~\ref{eq:filter3}).  We can determine all four parameters self-consistently, a situation known as \textit{extended} RLS.  

Unfortunately, the extended-RLS algorithm can diverge.  To make the algorithm converge, we use nominal $c_\pm$ values initially to estimate $\mu$ and $V_0$.  Then, after the initial estimates for $\mu$ and $V_0$ have stabilized,  we use those values to refine $c_\pm$ and again estimate $\mu$ and $V_0$.  To check that the extended-RLS algorithm converges  to the correct values, we have performed two tests:
\begin{itemize}
\item We  simulated time series and confirmed that the inferred parameter values for all four parameters were consistent with the simulation values.  (See Section~\ref{sec:simulations}.)
\item We independently measured the observation noise  directly, using the variance in the apparent position of an immobilized bead on a glass surface.\cite{Gavrilov2013}   The ``stuck bead" values of $\chi^2$ agreed with those found for diffusing particles using the extended-RLS algorithm.
\end{itemize}

\section{Two-dimensional feedback trap}
\label{sec:Trap2d}

For simplicity, the above discussion was for one dimension (1D), while the actual experiment explores two-dimensional (2D) motion, with the particle confined in the $z$ direction by using a thin cell.  The lateral $x$-$y$ coordinate system is defined in terms of the camera's pixel array.  Most of the previous discussion then directly generalizes to two dimensions.  Since the applied electric fields are not along the camera coordinate axes, the $x$-$y$ equations of motion are coupled and must be unscrambled.

Two sets of electrodes, Pairs 1 and 2, impose a 2D virtual potential.  The schematic diagram of the setup is given in Ref.~\onlinecite{Gavrilov2013}.  Empirically, the fields from Pairs 1 and 2 differ by up to $60 \%$ in magnitude and deviate from the $x$ and $y$ axes by up to  $45^\circ$.  We account for these effects by introducing a mobility matrix $\boldsymbol{\mu}$
\begin{equation}
	\boldsymbol{\mu} = 
	\begin{pmatrix} \mu_{x1} & \mu_{x2} \\ 
		\mu_{y1} & \mu_{y2} \end{pmatrix} \,,
\label{eq:mob12D}
\end{equation} 
where the subscripts indicate the transformation between $V_1$ and $V_2$ to $x$ and $y$ displacements.  Although $\boldsymbol{\mu}$ varies with position inside the cell (which measures 2 mm square), we find it to be constant over the scale of virtual potentials (several microns) at a fixed position within the cell.  We note that $\boldsymbol{\mu}$ is proportional to the mobility $\times t_s$, with a geometrical factor relating applied potentials at the electrodes to fields at the particle that must be calibrated empirically.

The 2D version of Eq.~\ref{eq:summary} then is  
\begin{equation}
	\bar{\boldsymbol{x}}_{n+1} = \bar{\mathbf{x}}_n + t_s\boldsymbol{\mu}(\bar{\boldsymbol{V}}_{n-1}-\boldsymbol{V}_{0}) + \boldsymbol{\zeta}_n \,,
\label{eq:LangevinVoltvector}
\end{equation} 
where all bold quantities other than $\boldsymbol{\mu}$ are 2D vectors.  The rest of the 1D analysis carries forward exactly as before, leading to filtered displacement equations of the form 
\begin{equation}
	\Delta \boldsymbol{\bar{x}}^{(f)}_n 
	= [\boldsymbol{\varphi}^{(f)}_n]^T \boldsymbol{\theta} + \boldsymbol{\psi}_n \,,
\label{eq:Filtered-fit2D}
\end{equation}
where $\boldsymbol{\theta}$ now has 6 elements (4 from the matrix $\boldsymbol{\mu}$ and 2 from $\boldsymbol{V_0}$).  The voltages are collected into a $2 \times 6$ matrix, which has only 2 independent components, made from the voltages from electrode pairs 1 and 2.  The noise $\boldsymbol{\psi}_n \sim \mathcal{N}(\boldsymbol{0},\boldsymbol{1})$, where $\boldsymbol{0} = \bigl( \begin{smallmatrix} 0 \\ 0 \end{smallmatrix} \bigr)$ and $\boldsymbol{1} = \bigl( \begin{smallmatrix} 1 && 0 \\ 0 && 1 \end{smallmatrix} \bigr)$.

Rather than writing out Eq.~\ref{eq:Filtered-fit2D} in components, it is more convenient to split it into two uncoupled equations with two individual RLS updates that are each based on three parameters, $\boldsymbol{\theta}_x^T = t_s \bigl( \begin{smallmatrix} \mu_{x1} && \mu_{x2} && \mu_{x1}V^{(1)}_0 + \mu_{x2}V^{(2)}_0 \end{smallmatrix} \bigr)$ and $\boldsymbol{\theta}_y^T =  t_s\bigl( \begin{smallmatrix} \mu_{y1} && \mu_{y2} && \mu_{y1}V^{(1)}_0 + \mu_{y2}V^{(2)}_0 \end{smallmatrix} \bigr)$.  These equations are
\begin{align}
	\overline{\Delta x}_n^{(f)} &= t_s
	\begin{pmatrix} \bar{V}^{(1)}_n & \bar{V}^{(2)}_n & 1 \end{pmatrix}^{(f)}
	\begin{pmatrix} \mu_{x1} \\ \mu_{x2} \\ \mu_{x1}V^{(1)}_0\!+\!\mu_{x2}V^{(2)}_0 \end{pmatrix} + \psi^{(x)}_n 
		\nonumber \\[3pt]
	\Delta \bar{y}_n^{(f)} &= t_s
	\begin{pmatrix} \bar{V}^{(1)}_n & \bar{V}^{(2)}_n & 1 \end{pmatrix}^{(f)}
	\begin{pmatrix} \mu_{y1} \\ \mu_{y2} \\ \mu_{y1}V^{(1)}_0\!+\!\mu_{y2}V^{(2)}_0 \end{pmatrix} + \psi^{(y)}_n  \,.	
\label{eq:Filtered-fit2Db}
\end{align}

From the RLS fit algorithm,\cite{Astrom2008} we calculate at each time step two vectors of length 3 that contain time-varying estimates of each fit parameter.  We also calculate two $3 \times 3$ covariance matrices that give the uncertainties of the best parameter estimates.  We update the best estimate and covariance matrix at each time step.  Since the $3 \times 3$ covariance matrix depends only on the inputs, it is the same for both equations and thus calculated only once per time step.  The forgetting algorithm is used to estimate $\boldsymbol{\mu}_n$ and $\boldsymbol{V_0}_n$, as in the 1D case.  The running averages of the diagonal elements of the matrices $\langle \boldsymbol{\zeta}_n \, \boldsymbol{\zeta}_n^T\rangle$ and $\langle \boldsymbol{\zeta}_n \, \boldsymbol{\zeta}_{n-1}^T\rangle$ are used to estimate the vectors $\boldsymbol{D}_n$ and $\boldsymbol{\chi}_n$ via Eq.~\ref{eq:DandChi1}.   (Physically, of course, we expect $D$ and $\chi$ to be the same along $x$ and $y$.  Calculating the two components independently checks that we have correctly decoupled the dynamics.)

From the imposed 2D potential $U(x,y,t)$, we generate forces by taking the negative gradient of the potential, $\boldsymbol{F}_n\equiv-\nabla U(\bar{x}_n,\bar{y}_n,t_n)$.  As in Sec.~\ref{sec:volts1d}, we update the voltages at each time step as
\begin{equation}
	\label{eq:Voltage2D}
          \boldsymbol{V}_{n} = - \boldsymbol{\mu}^{-1}[\boldsymbol{D}^{-1} \nabla U(\boldsymbol{\bar{x}}_{n}, \, t_{n})]  + \boldsymbol{V}_{0}
\end{equation}
where $\boldsymbol{D} = \bigl( \begin{smallmatrix} D_x && 0 \\ 0 && D_y \end{smallmatrix} \bigr)$.
One small effect that we do not model is that the noise components along $x$ and $y$ have a small cross-correlation arising from the fact that the number of photons detected is the same for both axes.  We do not observe any effects traceable to this small correlation.

\section{Simulations}
\label{sec:simulations}

Simulations of the particle dynamics are useful in showing that the rather complicated extended-RLS algorithm we propose here actually works.  That is, we will show that we can simulate a data set with known parameters ($\boldsymbol{\mu}$, $\mathbf{V_0}$, $\mathbf{D}$, and $\boldsymbol{\chi}$) and recover their values accurately.

\subsection{RLS estimate}
\label{sec:RLSsim}
In Sec.~\ref{sec:RLS}, we introduce the extended recursive least squares fit, with its associated filtering to decorrelate the noise.  Here, we use simulations to test the RLS algorithm and show that it converges to the correct values of material parameters in a feedback trap.  The method proposed in Sec. \ref{sec:RLS} does not depend on the shape of potential or on how voltages are applied; rather, the only requirements are that the voltages must vary sufficiently, so that they are sufficiently \textit{persistent}, in the language of adaptive control.\cite{Astrom2008}  The greater the voltage range explored, the faster the algorithm converges.

\begin{figure}[ht!]
	 \includegraphics[width=6cm]{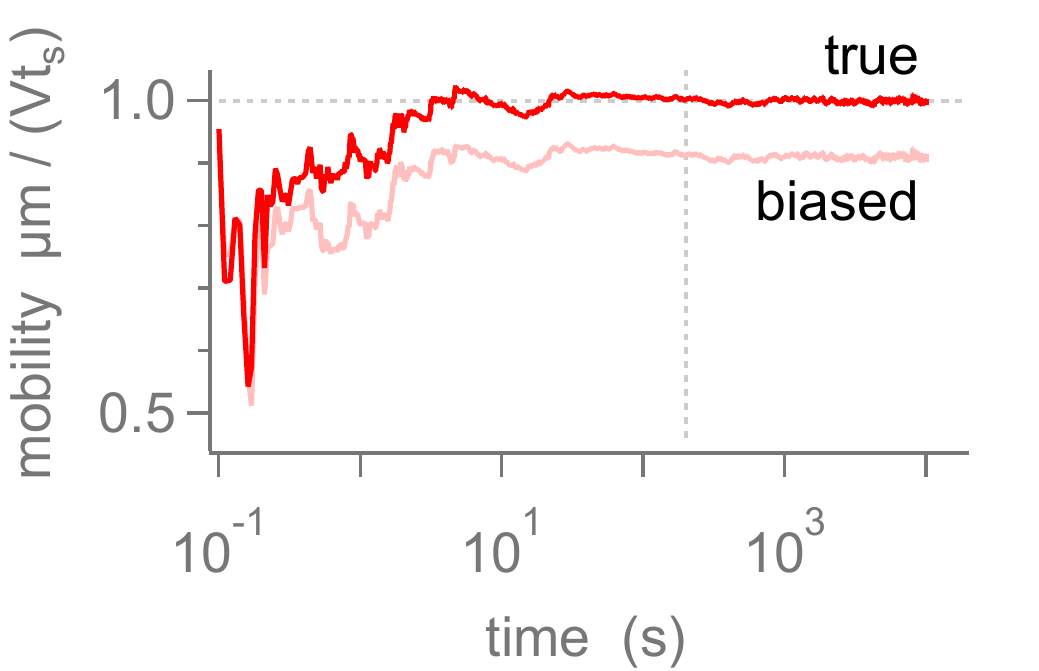}
	 \caption[example] { \label{fig:RLSsimulation} 
Simulation of RLS mobility estimate without (gray curve) and with (black curve) proper noise decorrelation.  Dashed line marks the mobility value used in the simulation, $\mu = 1$ $\mu$m / ($Vt_s$).  Vertical dashed line shows time used for calibration, 200 s.}
\end{figure}

In our experiments, we initially trap a particle in a virtual harmonic potential, to let the estimates of its material parameters (mobility and diffusion constant) converge sufficiently.  To check the procedure, we simulate a particle in the potential $U=\frac{1}{2}\kappa x^2$ and estimate its properties.  In Fig.~\ref{fig:RLSsimulation}, we show both filtered and naive estimates of the unknown mobility.  The naive RLS estimate, which neglects nearest-neighbor noise correlations, is biased down by $\approx 9\%$.  

We checked parameter convergence in two stages.  First, we assumed that we knew in advance both the diffusion constant $D$ and the observation noise $\chi$.  The latter can be pre-calibrated accurately, but the former varies from particle to particle---by a large amount, if the particle is an aggregate and by a small amount if an ``elementary" size.  But the first stage of parameter estimation requires a nominal value of $D$.  In a second set of simulations, we checked whether a bad guess affected parameter convergence and saw no difference, even when the initial estimate for $D$ was off by a factor of ten.

\section{Experimental data}
\label{sec:Experiment}
In the previous sections, we developed theoretical methods for measuring particle  properties in a feedback trap.  The framework handles properly (decorrelates) the different sources of noise in the feedback trap and was tested using simulations in Sec.~\ref{sec:RLSsim}.  Here, we show typical experimental data confirming that the various parameters do converge in practical settings.

\begin{figure}[ht!]
	 \includegraphics[width=6cm]{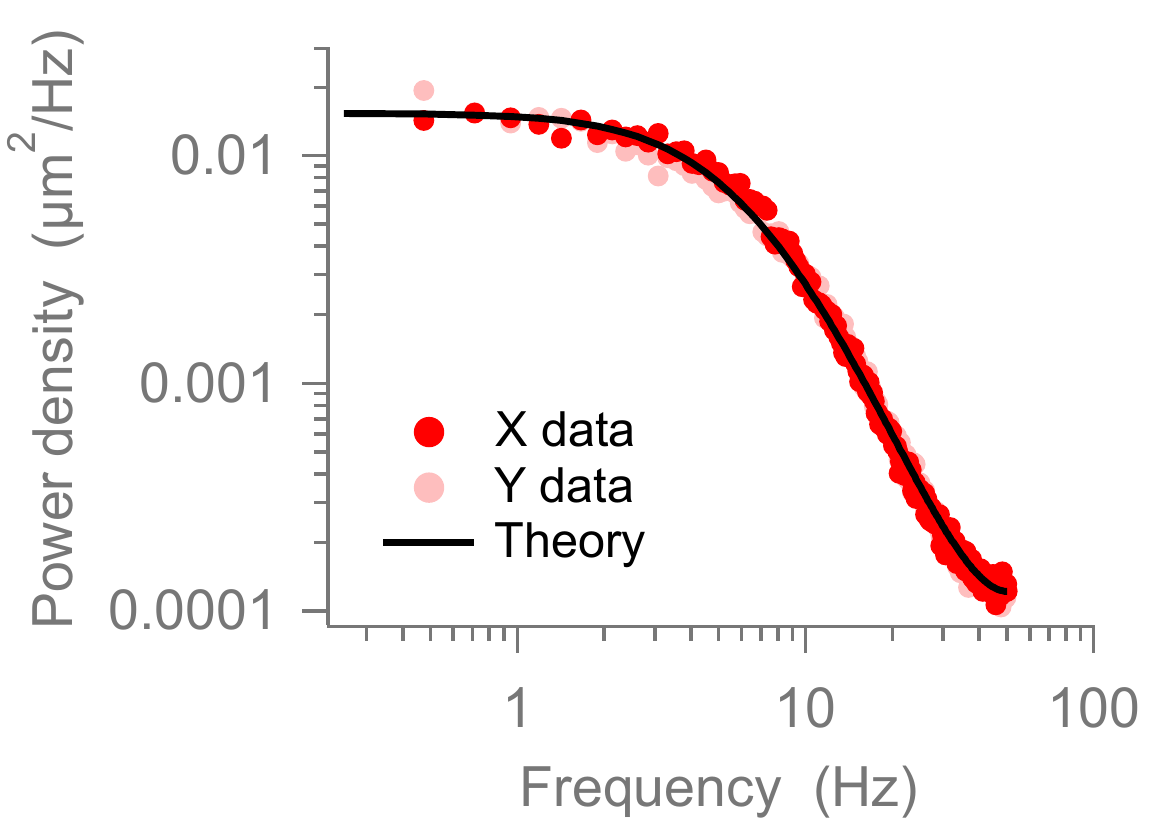}
	 \caption[example] { \label{fig:PSDexp} 
Power spectra for a particle in a feedback trap.  Dark (light) red markers indicate position measurements along the $x$- ($y$)-axis.  Solid black line shows the prediction from Ref.~\onlinecite{Jun2012}, plotted (not fit) using the independently inferred parameters.}
\end{figure}

We discussed a preliminary version of our experimental setup in Ref.~\onlinecite{Gavrilov2013} and gave first measurements of power spectrum and variance, as inferred from time series of position measurements in a virtual harmonic potential.  In that work, we were not able to measure particle properties while trapping.   The imposed feedback gain (equivalent to mobility) did not, in fact, match that measured in post-experiment analysis.  Moreover, the feedback gain drifted in time and was not constant.  As a result, we had to treat both the average feedback gain and diffusion coefficient as free parameters in a curve fit.  With the development of the extended-RLS algorithm presented here, we no longer need to fit parameters to the power spectrum.  We show  that we can use the RLS algorithm to estimate the parameters and then simply plot the power spectrum based on those parameters.  Figure~\ref{fig:PSDexp} shows the remarkable agreement that we can now achieve.  In particular, we note that the solid line, calculated according to the theory in Ref.~\onlinecite{Jun2012}, is not a fit but rather a plot, based on parameters taken from the extended-RLS formalism presented here.  This agreement justifies the rather complicated extended-RLS analysis of the parameters, the results of which we now describe in detail.

\subsection{Extended RLS estimates}
\label{sec:RLSexp}

The experimental test of our implementation of the extended-RLS algorithm (Fig.~\ref{fig:PSDexp}) uses a harmonic virtual potential with feedback gain $\alpha' = 0.2$.  At each time step, we apply a force proportional to the observed position, where $\alpha'
$ is the proportionality coefficient, $F_n = -\alpha' \tfrac{1}{Dt_s}\bar{x}_n$.  For small feedback gains, $\alpha' \approx \alpha = t_r/t_s$, which was defined above.  For the special case of a virtual harmonic potential, Eq.~\ref{eq:Voltage2D} implies that output voltages are calculated as $\boldsymbol{V}_{n}=-\alpha' \boldsymbol{\mu}^{-1} \boldsymbol{\bar{x}_n}/t_s + \boldsymbol{V_0}$.  In particular, they do not explicitly depend on a particle's diffusion constant.  Nevertheless, the extended-RLS estimation algorithm for $\boldsymbol{\mu}^{-1}$ and $\boldsymbol{V_0}$ uses the diffusion constant as an input, and the constant also affects the power spectrum density calculated from the particle's position measurements.

\begin{figure}[ht!]
	 \includegraphics[width=8cm]{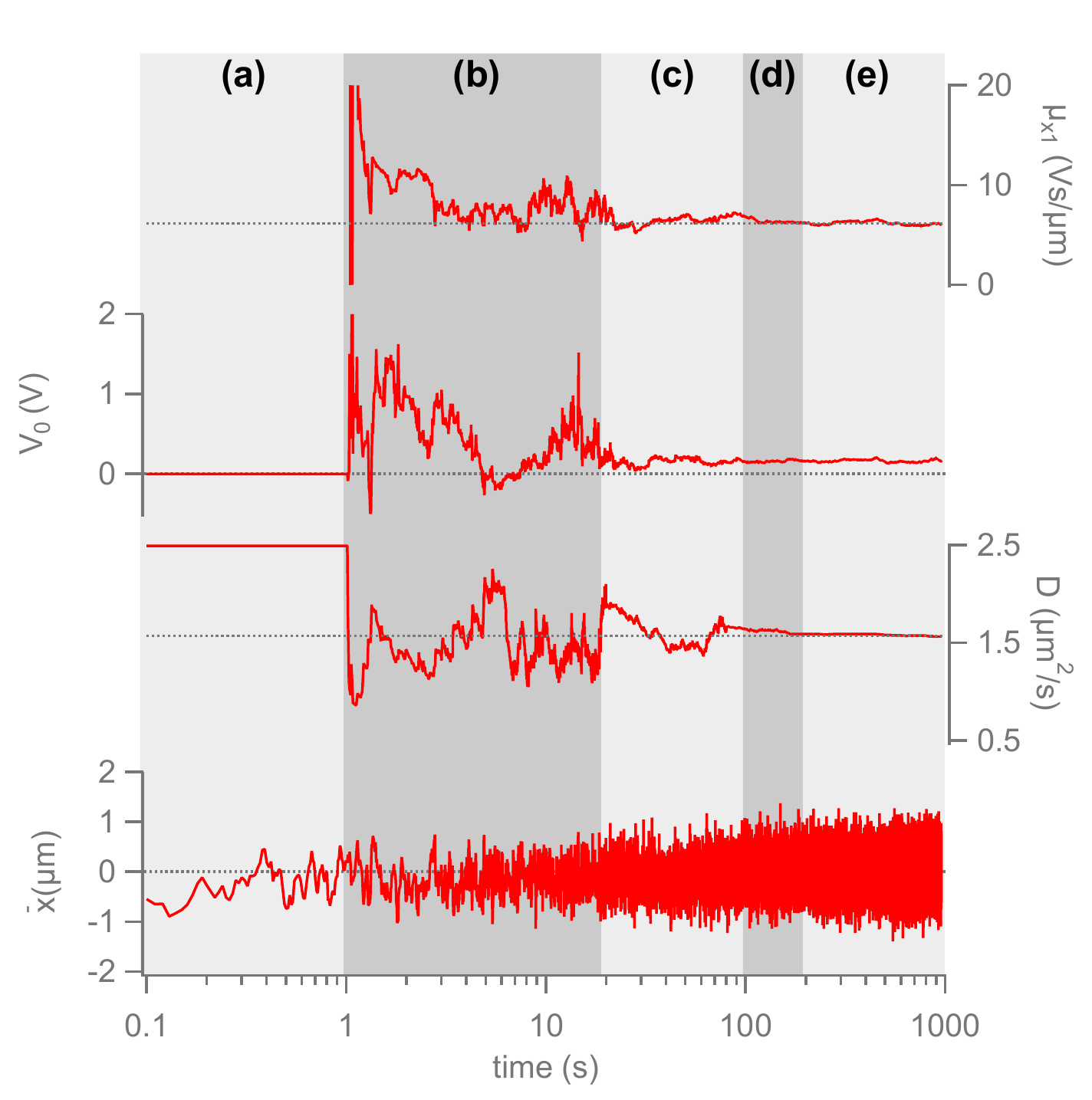}
	 \caption[example] {
Experimental estimates of the mobility component $\mu_{x1}$, drift voltage along one pair of electrodes $V_0$,  diffusion coefficient $D$, and observed position $\overline{x}$ along the $x$-axis.  Estimates of unknown parameters converge in 5 stages as the convergence time $\tau$ is adjusted, as discussed in App.~\ref{sec:details}:  (a)  No RLS estimate.  (b) and (c) Initial convergence with $\tau$ = 10 and $\tau$ = 100, respectively.  (d) Initial guesses for particle's properties are replaced by RLS estimates.  (e)  Full convergence, with time constant set to $\tau = 1000$.}
\label{fig:RLSexp} 
\end{figure}

As discussed in Appendix~\ref{sec:control}, we adjust the forgetting parameter $\lambda$ in several stages (in order to get the extended-RLS algorithm to converge).  Figure \ref{fig:RLSexp} shows an example of particle parameters recorded during an experiment run.  We estimate ten parameters:  four for the mobility $\boldsymbol{\mu}$, two for the drift $\boldsymbol{V_0}$, two for the diffusion $\boldsymbol{D}$ and two for the observational noise $\boldsymbol{\chi}$.    The time series in Fig.~\ref{fig:RLSexp} shows several stages in the convergence of  three parameters: the mobility component  $\mu_{x1}$, the drift voltage along one pair of electrodes $V_{0(1)}$, and the diffusion $D$ along $x$-axis.  The full convergence occurs over five stages, denoted \textbf{(a)}--\textbf{(e)}.  Stage \textbf{(a)} occurs after a new particle is detected. During this time, the illumination is adjusted so that the light intensity detected from the trapped bead matches the setpoint value.  Feedback voltages are generated using initial guesses for inverse mobility and drift.  In Stage \textbf{(b)}, the RLS estimate is turned on, with a short time constant $\tau=10$, which is increased to $\tau=100$ in Stage \textbf{(c)}.  (The time constants are given in units of $t_s = 10$ ms.)  The initial guesses are replaced by their  RLS estimates in Stage \textbf{(d)}.  Finally, in Stage \textbf{(e)}, the time constant is set to  $\tau=1000$, and we record data for further analysis.

Returning to Fig.~\ref{fig:PSDexp}, we see that the good agreement between power-spectrum data and the corresponding theory from \cite{Jun2012} requires accurate estimates of the timing parameters $t_s$ = 10 ms, $t_d$ = 10 ms (delay), and $t_c$ = 5 ms, of the observation noise $\chi$ = 40 nm, and of the diffusion constant $D = 1.54 \pm 0.06 \, \mu$m$^2$/s.  The  diffusion constant is calculated from the mean value of the RLS estimate for $D$ shown in Fig.~\ref{fig:RLSexp}, and its uncertainty is a systematic error due to the length calibration.  As discussed below, the statistical error is negligible.

\begin{figure}[ht!]
	 \includegraphics[width=8cm]{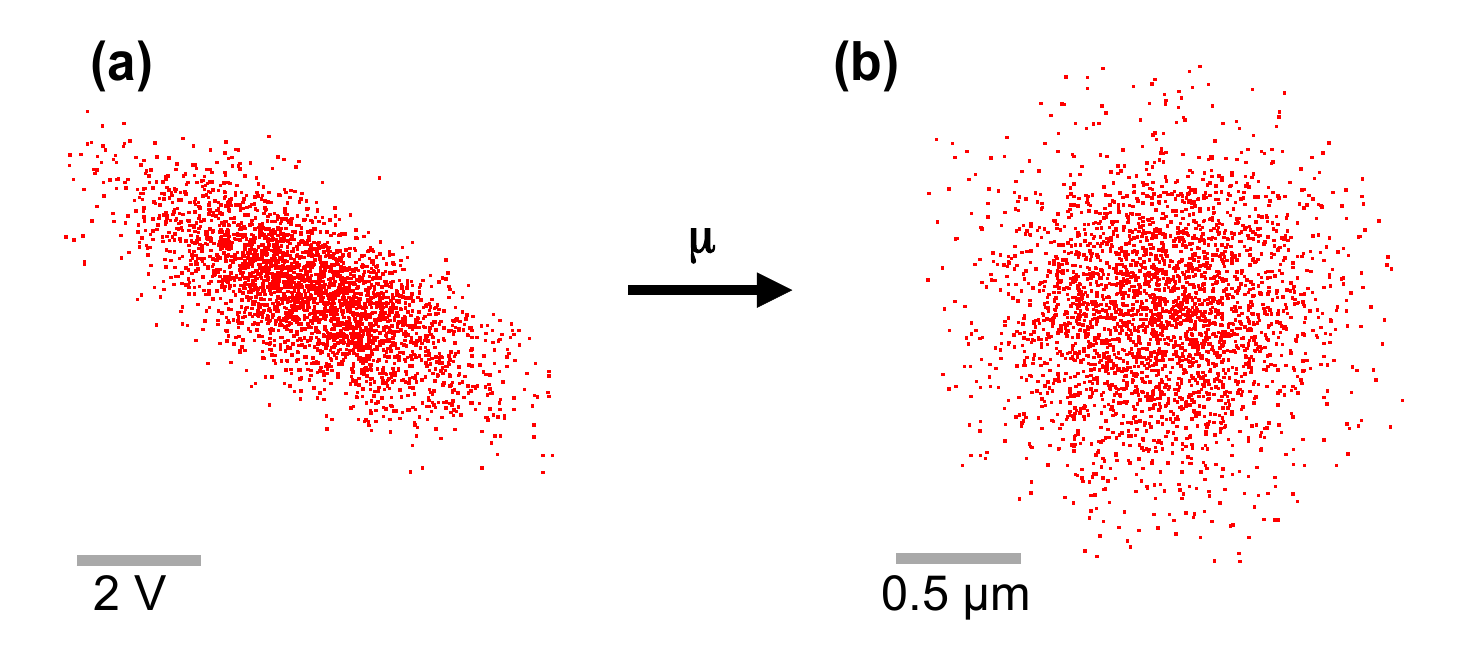}
	 \caption[example] { \label{fig:Density} 
Experimental scatter plot showing decorrelation transformation (the matrix $\boldsymbol{\mu}$ defined in Eq.~\ref{eq:mob12D}) between voltages and positions.  (a) Voltage applied to Electrode 2 vs. voltage applied to Electrode 1; (b) Position measurements, $y$ vs. $x$.}
\end{figure}

Figure~\ref{fig:PSDexp} also shows that the experimental data along both axes are similar, as expected physically (both directions are equivalent).  The result has stronger implications, as it also means that we have properly estimated the off-diagonal elements of the mobility matrix.  Incorrect values would lead to differences in the power spectrum, as well as cross correlations (arising because a voltage that is supposedly aligned along one camera axis has components along the other axis).  In Fig.~\ref{fig:Density}, we show how the applied voltages affects the observed position.  Although the applied voltages are highly correlated [Fig.~\ref{fig:Density} (a)], due to mobility matrix $\boldsymbol{\mu}$, they create an independent and uncorrelated position measurements [Fig.~\ref{fig:Density} (b)].  We use Fig.~\ref{fig:Density}(b), together with the power spectrum analysis in Fig.~\ref{fig:PSDexp}, to test whether particle dynamics follow the imposed virtual potential.

\subsection{Estimating the diffusion constant}
\label{sec:diffusion}

The extended RLS estimation algorithm successfully converges to the various parameter values used in numerical simulations and experimental runs.  But are the values that we deduce in the experiment correct?  To answer this question, we must first estimate independently the expected values of these parameters.  For the mobility, the charge on each particle is unknown, and the complexity of the electrode and cell geometry means that we do not know the electric field at the particle---only the voltage at the electrodes.  Mobility measurements are thus relative.

We can do better estimating the lateral diffusion constant.  For an isolated sphere of radius $r$ in a fluid of viscosity $\eta$ of infinite extent, the Stokes-Einstein relation gives $D_\infty = kT/\gamma$, with the fluid drag coefficient $\gamma = 6\pi r \eta$.  However, $r$, $T$, and $\eta$ must be estimated.  In addition, the particle is not in an infinite fluid medium but is confined between two parallel plates, an effect that increases the drag $\gamma$ and reduces the measured $D$.

\subsubsection{Diffusion in an infinite medium}

We first estimate $D_\infty$.  The radius is known to $\pm 3\%$.\cite{Gavrilov2013}  The temperature varies considerably day to day in the laboratory but can be measured to $\pm 1^\circ$C by placing a thermistor near the sample during the experiment.  The uncertainty in absolute temperature is small $(0.3\%)$, but that of water viscosity due to temperature uncertainty is more significant, $\pm 3\%$.\cite{HandbookCH2012}  Together, these imply a $\pm 5\%$ uncertainty in $D_\infty$.  For $r=100$ nm and $T=26^\circ$C, the nominal value is 2.4 $\pm$ 0.1 $\mu$m$^2$/s.

\subsubsection{Diffusion in a confined medium}

The dominant uncertainty in the estimation of $\gamma$ and $D$ is due to the uncertainty in the absolute sample thickness.  Because the particle diameter is $\approx 210$ nm and the nominal plate spacing is $\approx 800$ nm, confinement effects are strong.  Qualitatively, the extra shear between fluid that pinned to the moving sphere boundary and fluid at the stationary cell plate boundary increase $\gamma$ and reduce $D$.  Below, we argue that the electrostatic repulsion from the walls implies that the particle is usually near the center of the cell, where the diffusion coefficient $D_{\rm mid}$ is well approximated by the fifth-order expansion,\cite{happel1983}
\begin{equation}
	\frac{D_{\rm mid}}{D_\infty} \approx 1-1.004\beta
		+0.418\beta^3+0.21\beta^4-0.169\beta^5 \,,
\label{eq:ConfDif}
\end{equation}
where $\beta = r/d$ is the ratio between the particle radius $r$ and the cell thickness $d$.  Unfortunately, while we use a profilometer to measure cell thickness prior to assembly, we do not have a direct way of measuring $d$, the thickness of the assembled, filled cell.  We estimate the uncertainty in $d$ to be about $\pm 200$ nm ($\pm$ 25\%) from the optical images of particles that move in and out of focus in the feedback trap.  The depth of focus is $\approx 500$ nm, and we can compare particle images to standard out-of-focus images, leading to a range of 600--1000 nm in particular samples and a similar $\pm 25\%$ uncertainty in $D$.

\subsubsection{Electrostatic repulsion from the walls}

We have assumed that the particle is mostly near the cell's midplane.  How good is this approximation?  To avoid having particles stick to the cell walls, we ensure that both the glass walls and latex spheres are negatively charged.\cite{Gavrilov2013}  Electrostatic forces in water are screened by counterions, over a length scale quantified by the Debye length ($\lambda_D$).  The Debye length in pure, deionized water is 960 nm, but CO$_2$ in the air and impurities from the sample cell reduce the screening.  Behrens and Grier \cite{Behrens2001} reported the Debye length in ``real" water to be 275 nm.  Although much reduced, such a length is still significant on the scale of the 800 nm cell and implies that the particle will spend most of its time near the midplane of the cell. 

\begin{figure}[ht]
	 \includegraphics[width=6cm]{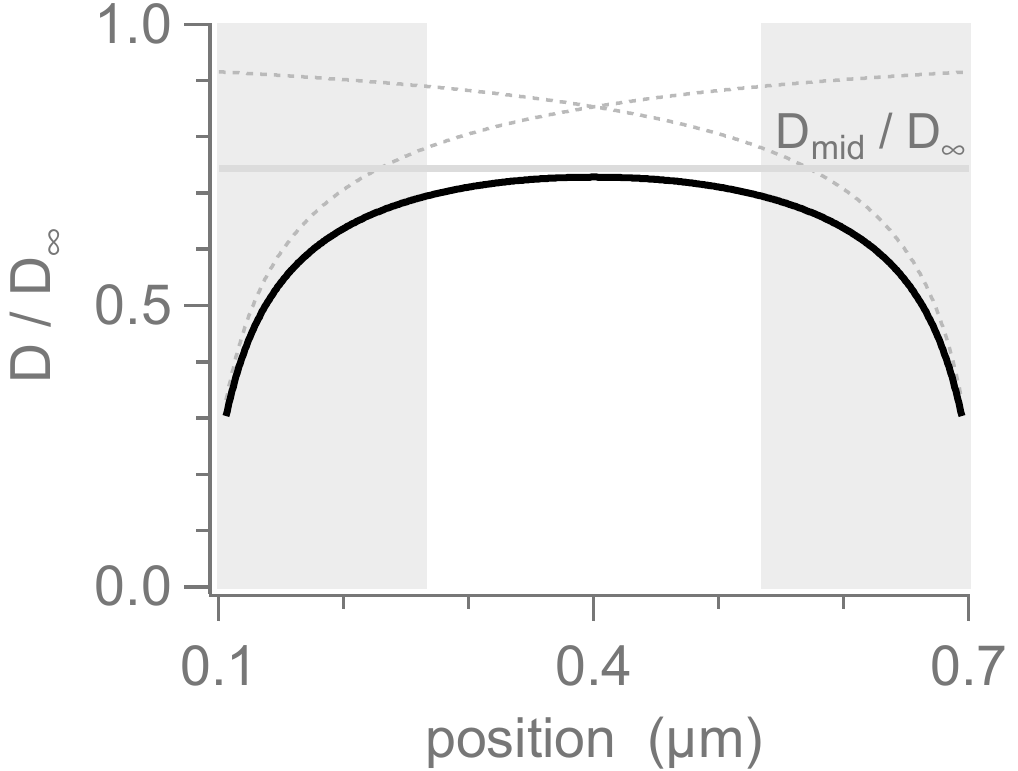}
	  \caption[example] { \label{fig:Debye}  Effects on the diffusion in confined geometry due to hydrodynamic and electrostatic effect for a cell of thickness $d=800$ nm.  The solid line is based on the superposition of two independent one-wall corrections (dotted gray lines).  The solid gray line is the midpoint diffusion approximation. Shaded area is the Debye length $\lambda_D$.}  
\end{figure}

Figure \ref{fig:Debye} combines our analysis of hydrodynamic drag and electrostatic repulsion in a confined geometry.  The gray shaded area shows the screening (Debye) length, from which the particle is effectively excluded.  The particle is nearly always in the central area, where $D \approx D_{\rm mid}$, with at most a 5\% overestimate.  

\subsubsection{Test of diffusion measurements}

As a quantitative test of these ideas, we measured $D$ in a cell with nominal parameters $d = 800$ nm, $T= 26 ^\circ$C, $r=105$ nm.  Using 400 s of data in a harmonic trap, we found $D = 1.54 \pm 0.06$ $\mu$m$^2$/s, where the uncertainty is dominated by the precision of the length calibration between camera pixels and absolute length.  (The statistical error from the fit, $4 \times 10^{-5}~\mu$m$^2$/s, is negligible.)  The experimental measurement is consistent with the mid-plane estimate of $D_{\rm mid} = 1.8 \pm 0.5$ $\mu$m$^2$/s.  The uncertainty in the latter estimate is dominated by the systematic error due to the uncertainty in cell thickness.   Using a smaller particle or thicker cell would reduce that source of systematic error.  Such tests are not possible in the current setup, as smaller particles require faster update times $t_s$, and thicker cells require either tracking the vertical direction or an imaging system with large depth of focus.  Nonetheless, the important point is that the values that we measure for the diffusion constant are consistent with expectations, given the experimental geometry.

\section{Conclusion}
\label{sec:conclusion}

We have shown that an extended-RLS algorithm allows one to reconstruct accurately the mobility and diffusion constants of particles in a feedback trap.  In further work, we have used the procedures developed here to study Landauer's principle, which gives a lower bound to the amount of work needed to erase a bit of information.

It is interesting to compare the present study with the recent work of Wang and Moerner\cite{Wang2014}.  In the latter work, camera images are replaced by a continuously scanning laser and a single-pixel detector that detects individual photon counts.\cite{jiang08,Fields2011}  Each individual count leads to an updated estimate of particle position, using a modified Kalman filter.  The advantage of such a setup is vastly increased speed and a simplicity that comes from having instantaneous estimates of particle position (no camera exposure effects).  The limitations are that interpreting extended objects becomes difficult.  In addition, the presence of background photons leads to non-Gaussian statistics and the need for more sophisticated algorithms (expectation-maximization in Ref.~\onlinecite{Wang2014} and assumed density filter in a related work\cite{fields12}).  Both of those algorithms are significantly more complicated than the extended-RLS algorithm here.  

The algorithm given here will simplify even more if we can shorten the camera exposure to make the camera-correction terms not just small but completely negligible.  To keep the observation noise at the same level, we would then have to compensate for the shorter exposure by increasing the illumination intensity.  Higher intensities can lead to accelerated photobleaching of fluorescent molecules, but non-fluorescent imaging that depends on detecting scattered light would not suffer from such problems.  Gold nanoparticles, which show strong plasmonic scattering at wavelengths that depend on their size, are attractive candidates for such studies.

\begin{acknowledgments}

This work was funded by NSERC (Canada).  The microfluidic flow cell was fabricated in the 4D LABS facility at Simon Fraser University.

\end{acknowledgments}

\appendix

\section{Effect of camera exposure}
\label{sec:camera-exposure}
To see that $\overline{F}_{n-1} \approx F_{n-1}$, we define the deviation $\delta F_{n-1} = \overline{F}_{n-1}-F_{n-1}$ and consider
\begin{align}
	\frac{\langle(\delta F_{n-1})^2\rangle}{\langle F^2 \rangle} 
	&= \left( \frac{t_c}{8t_s} \right)^2  \frac{ \langle (F_n-2F_{n-1}+F_{n-2})^2 \rangle}
		{\langle F^2 \rangle} \nonumber \\[3pt]
	&= \left( \frac{t_c}{8t_s} \right)^2 \frac{ \left(6 \langle F^2 \rangle 
		- 8 \langle F \, F_{-1} \rangle + 2 \langle F \, F_{-2} \rangle \right)}
		{\langle F^2 \rangle}  \,,
\label{eq:force-correction-size}
\end{align}
where $\langle F^2 \rangle \equiv \langle F_n^2 \rangle$.  We drop the $n$ index because of the homogeneity in time.  The $\langle F \, F_{-1} \rangle$ term similarly contains contributions from both $\langle F_{n} \, F_{n-1} \rangle$ and $\langle F_{n-1} \, F_{n-2} \rangle$.  Next, we recognize that the forces $F_n$ are correlated on a relaxation time scale $t_r$ that is given by $t_r = \kappa/\gamma$, where $\gamma$ is the fluid drag and where $\kappa$ is the local curvature of the potential.  That is, near the position $x_n$, the potential is locally $U(x) \approx \tfrac{1}{2}\kappa(x-x_n)^2$.  If we define, as above, $\alpha = t_s/t_r$, then the overdamped-dynamics correlations are given by
\begin{equation}
	\langle F \, F_{-p} \rangle \approx \langle F^2 \rangle e^{-|p|\alpha} \,,
\label{eq:force-corr}
\end{equation}
The expression in Eq.~\ref{eq:force-correction-size} is then
\begin{align}
	&\quad \, 6 \langle F^2 \rangle - 8 \langle F \, F_{-1} \rangle + 2 \langle F \, F_{-2} \rangle
		\nonumber \\ 
	&\approx 6 \langle F^2 \rangle - 8 \langle F^2 \rangle e^{-\alpha} 
		+ 2 \langle F^2 \rangle e^{-2\alpha} \nonumber \\
	&= 4\alpha \langle F^2 \rangle + \mathcal{O}(\alpha^2) \,.
\end{align}
Substituting back into Eq.~\ref{eq:force-correction-size} then gives
\begin{equation}
	\sqrt{\frac{\langle(\delta F_{n-1})^2\rangle}{\langle F^2 \rangle}}
	= \left( \frac{\sqrt{\alpha} t_c}{4t_s} \right) \approx 0.06 \,,
\label{eq:force-correction-size1}
\end{equation}
for $\alpha = 0.2$ and $t_c/t_s = 0.5$.  We note that $\bar{V}_n \approx V_n$, by the same argument.

\section{Control program}
\label{sec:control}

As noted in the main text, the extended-RLS algorithm can diverge, and careful attention to the initial iterations is crucial for convergence.  These initial iterations inevitably are strongly coupled to the actual functioning of the control program.  The experimental protocol is complicated, because particles may bleach and need to be discarded, because new unwanted particles may diffuse into the field of view, etc.  Here, we first give a simplified overview of trap operation and then provide a more detailed account.

\subsection{Overview}
\label{sec:overview}

Figure~\ref{fig:FlowChartSimple} shows a simplified flowchart of the control software.  The basic structure is that of a \textit{state machine}:  at each time step, an image is acquired, the program determines the trap state and then responds by selecting a given operational mode.  In the simplified version given in Fig.~\ref{fig:FlowChartSimple}, there are three modes:

\begin{figure}[h]
	 \includegraphics[width=8cm]{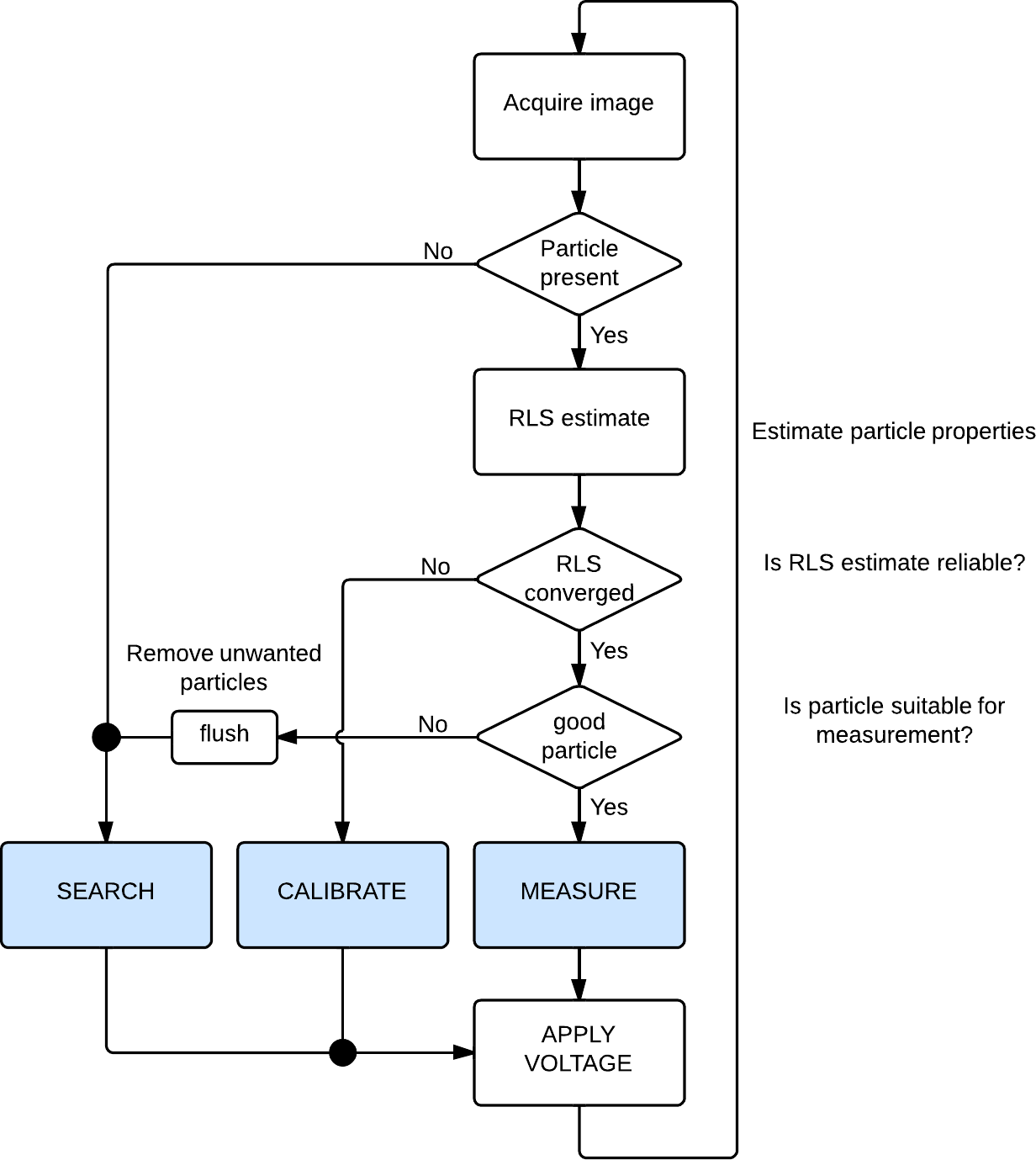}
	 \caption[example] { \label{fig:FlowChartSimple} 
Simplified flowchart of the control program.  The three basic modes of operation are indicated by the shaded blue boxes.}
\end{figure} 

\begin{itemize}
\item \textit{Search}:  No particle is present and we apply a large voltage to sweep new particles into the field of view.

\item \textit{Calibrate}:  A new particle is present and its properties are being calibrated, a process that takes 400 s.

\item \textit{Measure}:  A calibrated particle is present, and we apply the appropriate virtual potential, updating the calibration to account for drifts.  The update uses the extended RLS algorithm described above.

\end{itemize}

In the simplified Fig.~\ref{fig:FlowChartSimple}, the trap operates mostly in the \textit{Measure} state and follows a loop that starts from ``Acquire image" and proceeds downwards to ``Apply voltage," before looping back to acquire another image.  If there are problems---no good particle is detected (i.e., because it has bleached or because an unwanted particle has diffused into the field of view)---then the system switches to search mode, where it flushes the existing particle and searches for a new one by imposing a large, constant voltage.  Once a new particle has been found, the trap enters a calibration mode, to let parameter estimates stabilize, before new measurements can proceed.

\subsection{Details}
\label{sec:details}

\begin{figure}[ht!]
	 \includegraphics[width=8cm]{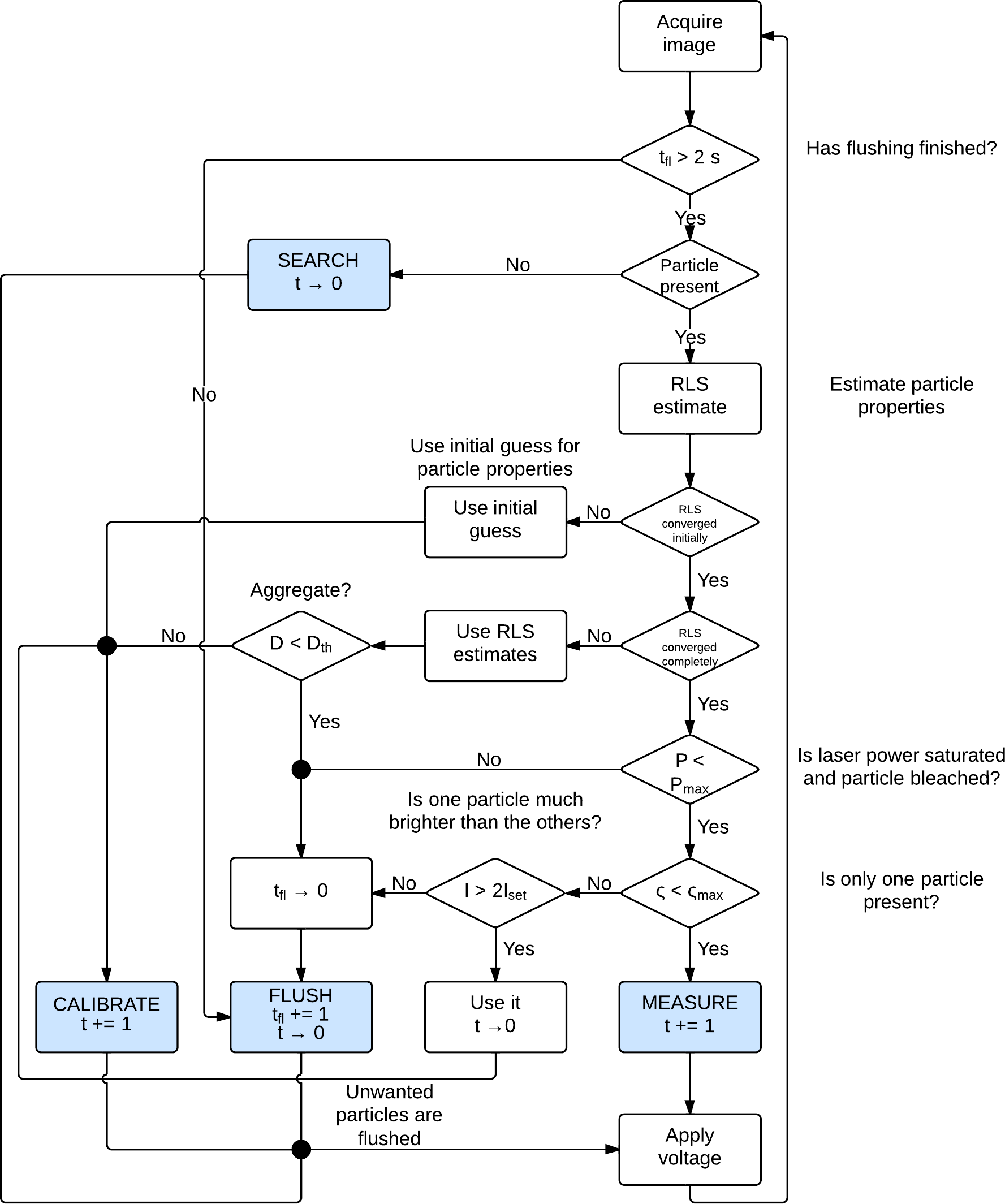}
	 \caption[example] { \label{fig:FlowChartComplex} 
Full flow chart with automated event identification and detailed event handling.}
\end{figure}

The description in Section~\ref{sec:overview} leaves out many details.  Figure~\ref{fig:FlowChartComplex} shows the complete flowchart for trap operation.  There are eight possible particle states, and the program responds by operating in one of four modes of operation.  The states depend on the number of particles in the field of view (0 or 1 or $\ge 2$) and are further classified into substates, as follows: 
\begin{itemize}
\item \textit{Single particle}:  good $|$ unknown $|$ partially known $|$ bleached $|$ aggregate
\item \textit{Two or more particles}:  comparable intensities $|$ one is significantly brighter
\item \textit{No particle present}.
\end{itemize}
The modes of operation are \textit{Search, Calibrate, Measure, Flush}.

We also introduce two different timers. The first is denoted by $t$ and measures the time since a new particle was detected.  This time is also used for defining the protocols of time-dependent potentials.  The second time is denoted by $t_{\rm fl}$ and counts the  time since an unwanted particle was detected.  It is reset to zero when an unwanted particle is detected and when unwanted particles are flushed.  Flushing is accomplished by setting a large voltage (6 V).  The mode operates until $t_{\rm fl} =$ 2 s.  Note that images are acquired every $t_s = 10$ ms while the trap operates in flush mode, but their content is ignored until the flush is complete.

After the cell is flushed, we maintain the same large DC voltage.  Now, however, its purpose is to search for a new particle.  We then analyze each acquired image.  Based on the summed light intensity in the image, we determine whether a particle is present, \cite{Gavrilov2013} which we infer if the intensity is higher than a threshold value ($I>I_{th}$) during two consecutive time steps.  We input the observed position of a new particle and the associated applied voltages into the RLS algorithm to estimate the material parameters.  We occasionally encounter instability when the initial parameter guess is too far off, especially when they are over estimated.  In such cases, the program simply flushes and then reverts to search mode.

When a new particle is detected, we reset the covariance matrix $\boldsymbol{P}$ in Eq.~\ref{eq:RLS-time-var} by multiplying all elements by $10^4$.  We also set the forgetting factor to $\lambda$ = 0.99 (or $\tau$ = 100).  Then, after 20 s, we increase $\tau$ to 1000 and continue to update the RLS algorithm.  For $100 < t <  200$~s, we use an ordinary RLS algorithm with fixed $c_\pm$ to estimate $\boldsymbol{\mu}$ and $\boldsymbol{V}_0$.  During this time, the RLS algorithm converges to a constrained steady state.  For $t > 200$ s, we let the noise parameters $c_\pm$ vary in the full extended RLS algorithm.  This elaborate initialization procedure keeps the extended-RLS algorithm from diverging.

After the initial RLS convergence ($t=200$ s), we increase the forgetting factor lifetime to $\tau =10 \, 000$ and start the full extended RLS algorithm.  We also start to estimate $\boldsymbol{D}$ and $\boldsymbol{\chi}$, using Eq.~\ref{eq:DandChi1}.  The RLS convergence is not sensitive to the value of $\boldsymbol{D}$; however, the shape of imposed potential is.  It takes an additional $200$~s for the RLS algorithm and $\boldsymbol{D}$ estimate to fully converge ($t < 400$~s).  If the estimated diffusion coefficient is smaller than a threshold value  $D_{\rm th}$, we conclude that the particle is an aggregate and flush it. 

After the RLS algorithm has fully converged, we impose the virtual potential and perform the work measurements.  At each time step, we check whether a particle has bleached, by examining the laser output power.  The laser operates in a proportional-integral feedback loop that attempts to keep the fluorescence intensity constant by altering the input laser power, which saturates at a high value when the particle is too dim. \cite{Gavrilov2013}  If the particle has bleached, we flush it.  We also test the noise term $\zeta_n$ at each time step.  From Eq.~\ref{eq:zeta_n}, this noise term can be interpreted as the difference between the measured displacement $\overline{\Delta x}_n$ and the displacement $ t_s\mu(\bar{V}_{n-1}-V_{0})$ imposed by the feedback trap.   If the difference between observed and imposed displacements is too big ($\zeta_n>\zeta_{\rm max}$), then two or more particles may be present in the observation area.  This happens when a new, unwanted particle diffuses into the observation area and then is mis-tracked by image analysis algorithm.  The inferred position of such a particle is usually located in between the actual positions of old and new particles, leading to a sudden, large ``displacement" $\zeta_n$.  In this case, we flush both particles.  Sometimes, one particle is significantly brighter than the other.  In this case, we keep it and use it to calibrate the system.  The dimmer particle is not trapped and quickly leaves the field of view.  It also becomes immediately less visible after the AOD control algorithm reduces the laser intensity.

Each of the four possible modes of operation generates two voltages at each time step, which are sent to the two electrode pairs to create the desired displacements.

\bibliography{cal}

\end{document}